\documentclass[useAMS]{mn2e}
\usepackage{epsfig,natbib_vw,times}

\usepackage{bm}

\bibpunct[, ]{(}{)}{;}{a}{}{,}

\def \aj {AJ}
\def \mnras {MNRAS}
\def \apj {ApJ}
\def \apjs {ApJS}
\def \apjl {ApJL}
\def \aap {A\&A}
\def \nat {Nature}

\def \pasp {PASP}

\def \be {\begin{equation}}
\def \ee {\end{equation}}

\def\gsim{\mathrel{\lower0.6ex\hbox{$\buildrel {\textstyle >}
 \over {\scriptstyle \sim}$}}}
\def\lsim{\mathrel{\lower0.6ex\hbox{$\buildrel {\textstyle <}
 \over {\scriptstyle \sim}$}}}

\def\m@th{\mathsurround=0pt }
\def\eqalign#1{\null\,\vcenter{\openup1\jot \m@th
 \ialign{\strut\hfil$\displaystyle{##}$&$\displaystyle{{}##}$\hfil
 \crcr#1\crcr}}\,}

\def \mum {$\mu$m}

\def \logM {$\log({\rm M}/{\rm M}_\odot)$}

\def \logphi {$\log(\phi/{\rm Mpc}^{-3})$}

\title[The evolution of post-starburst
  galaxies]{The evolution of post-starburst
  galaxies from \boldmath{$z=2$} to \boldmath{$z= 0.5$}}
\author[V. Wild et~al.]{
\parbox[t]{\textwidth}{\raggedright 
Vivienne Wild$^1$\thanks{vw8@st-andrews.ac.uk},  Omar Almaini$^2$, Jim
Dunlop$^3$, Chris Simpson,  Kate Rowlands$^1$ , Rebecca Bowler$^4$, David Maltby$^2$,
Ross McLure$^3$
}
\vspace*{6pt}\\
$^1$SUPA\thanks{Scottish Universities Physics Alliance}, School of
Physics and Astronomy, University of St Andrews, North Haugh, St Andrews, KY16 9SS, U.K. \\
$^2$University of Nottingham, School of Physics and Astronomy, Nottingham, NG7 2RD, U.K\\
$^3$SUPA$\dagger$, Institute for Astronomy, University of Edinburgh, Royal
Observatory, Blackford Hill, Edinburgh, EH9 3HJ, U.K. \\
$^{4}$Subdepartment of Astrophysics, University of Oxford, The Denys
Wilkinson Building, Keble Road, Oxford, OX1 3RH, U.K. 
}

\begin{document}
\maketitle
\begin{abstract}

  We present the evolution in the number density and stellar mass
  functions of photometrically selected post-starburst galaxies in the
  UKIDSS Deep Survey (UDS), with redshifts of $0.5<z<2$ and stellar
  masses \logM$>10$.  We find that this transitionary species of
  galaxy is rare at all redshifts, contributing $\sim$5\% of the total
  population at $z\sim2$, to $<$1\% by $z\sim0.5$. By comparing the
  mass functions of quiescent galaxies to post-starburst galaxies at
  three cosmic epochs, we show that rapid quenching of star formation
  can account for 100\% of quiescent galaxy formation, if the
  post-starburst spectral features are visible for
  $\sim$250\,Myr. The flattening of the low mass end of the quiescent
  galaxy stellar mass function seen at $z\sim1$ can be entirely
  explained by the addition of rapidly quenched galaxies. Only if a
  significant fraction of post-starburst galaxies have features that
  are visible for longer than 250\,Myr, or they acquire new gas and
  return to the star-forming sequence, can there be significant growth
  of the red sequence from a slower quenching route. The shape of the
  mass function of these transitory post-starburst galaxies resembles
  that of quiescent galaxies at $z\sim2$, with a preferred stellar
  mass of \logM$\sim10.6$, but evolves steadily to resemble that of
  star-forming galaxies at $z<1$. This leads us to propose a dual
  origin for post-starburst galaxies: (1) at $z\gsim2$ they are
  exclusively massive galaxies that have formed the bulk of their
  stars during a rapid assembly period, followed by complete quenching
  of further star formation; (2) at $z\lsim1$ they are caused by the
  rapid quenching of gas-rich star-forming galaxies, independent of
  stellar mass, possibly due to environment and/or gas-rich major
  mergers.

\end{abstract}


\section{Introduction}\label{sec:intro}

In the local Universe, the population of massive galaxies is bimodal
in both structure and spectral type: the majority can be described as
either being quiescent and elliptical, or star-forming and
spiral. Understanding the origin of this bimodality in the galaxy
population has been the topic of considerable research for many
decades. Recently, large multiwavelength surveys have allowed us to
measure the fraction of quiescent and star-forming galaxies over a
large portion of cosmic time, providing direct observations of the
emergence of galaxy bimodality. These have shown that both the number
density of quiescent galaxies, as well as their total stellar mass
density, increases steadily with time since at least $z\sim3-4$
\citep[e.g.][]{2004ApJ...608..752B,2007ApJ...665..265F,Ilbert:2013p9113,
  Muzzin:2013p9592}, although details depend on the survey as well as the
methods used to calculate photometric redshifts and stellar masses,
and to identify quiescent galaxies. As quiescent galaxies are unable
to form new stars in-situ, the increase in the global stellar mass
density of quiescent galaxies must arise from a cessation of star
formation in star-forming galaxies and transfer of galaxies from the
star-forming to quiescent populations. This process has come to be
known as galaxy ``quenching'', and one of the central questions in
galaxy evolution research is which physical processes are responsible
for this shut down in star formation in a portion of the galaxy
population.

The presence of quiescent galaxies out to redshifts of $z\sim4$ or
more \citep[e.g.][]{Straatman:2014p9600}, shows that star formation
can be rapidly quenched following an early formation epoch. These
early quiescent galaxies are found to be elliptical and highly compact
\citep[e.g.][]{Daddi:2005p9602,Damjanov:2009p9608,vanderWel:2014p9612}.
The current generation of simulations suggest that early quiescent,
elliptical systems could have formed through strongly dissipative
processes such as disk instabilities and mergers, where there is a
rapid gas inflow towards the centres of the galaxies
\citep{Hopkins:2009p10121,Wuyts:2010p10007,Dekel:2014p9459,Zolotov:2015p9460,Ceverino:2015p10008}.  In these simulations, massive and efficient outflows are also
required to expel the remainder of the gas that is not consumed by
star formation.  This feedback may be caused by the rapid gas accretion
leading to self-regulated star formation via supernovae feedback
\citep[e.g.][]{Muratov:2015p9462}; ``quasar-mode'' active galactic
nuclei (AGN) have also been suggested to play a role in the final removal of gas from the
system \citep[e.g.][]{Silk:1998p2398,2005Natur.433..604D}. These
postulated feedback mechanisms should act quickly, over
timescales of order $10^8$ years and could explain the rapid quenching of
galaxies observed. Regardless of the mechanism
invoked to explain the initial quenching event, maintaining the quiescence of
high-redshift galaxies in such a gas rich environment remains a
challenge for models \citep[e.g. see the discussion
in][]{Gabor:2010p9903}, with scenarios invoking suppression of a fresh
gas supply by shocks maintaining a hot circum-galactic halo
\citep[e.g.][]{Dekel:2006p9464,Zolotov:2015p9460}, or ``radio-mode''
AGN feedback
\citep[e.g.][]{2006MNRAS.370..645B,2006MNRAS.365...11C}. Observationally,
\citet{Best:2006p9940} showed that radio-loud AGN could provide
sufficient energy to prevent the further growth of massive elliptical
galaxies, however, direct observational evidence for feedback
mechanisms to directly remove significant quantities of gas from
galaxies is largely lacking.

 At lower redshifts ($z\lsim 2$), the continued increase in the
  number density of quiescent galaxies
  \citep[e.g.][]{Moustakas:2013p8856,Moutard:2016p10257} shows that
  processes remain in place that quench star formation, even as the
  gas fractions in galaxies decrease, disks become more stable and
  cold gas streams shut down.  The relatively constant number density
  of star-forming galaxies, combined with the global decline in star
  formation rate density by a factor of $\sim10$ since $z\sim2$
  \citep{Madau:2014p10259}, indicates that typical star forming
  galaxies are gradually decreasing their star-formation rates over
  timescales of several Gyr. However, it is unclear to what extent this
  decrease in the average star formation rate of star-forming galaxies
  is related to the formation of completely quiescent galaxies. In
  particular, the approximately coeval increase in the fraction of
  galaxies with elliptical-like morphologies
  \citep[e.g.][]{Bruce:2012p8897,Buitrago:2013p9466}, and the strong
  correlation between quiescent galaxies and a predominantly spheroid
  morphology at all redshifts
  \citep[e.g.][]{Bell:2012p9465,Lang:2014p9468,Bruce:2014p9467,Brennan:2015p9469,Ownsworth:2016p10260},
  suggests that the dominant quenching mechanism is more than a simple
  depletion of gas supplies.

  Gas-rich major mergers have long been considered a strong candidate
  for forming at least some types of elliptical galaxies, and are
  perhaps particularly relevant at $z<1$ where the additional processes that can
  lead to dissipational collapse at high-redshift become less
  likely. Numerical simulations indicate that gas-rich mergers are
  able to shut down star formation through rapid depletion of gas
  supplies, potentially accompanied by strong feedback, as well as
  disrupt the stellar orbits leading to an elliptical remnant
  \citep[e.g.][]{1972ApJ...178..623T,Naab:2006p9458,Cox:2006p9454}. However,
  cosmological hydrodynamical simulations illustrate the many
  complexities of the problem \citep{Naab:2014p9455}, disks may
  survive or regrow following a merger
  \citep{2009MNRAS.398..312G,2009ApJ...691.1168H,2016ApJ...821...90A},
  and observationally it remains difficult to firmly link major
  mergers to the increasing quiescent fraction.  

Ultimately, the assembly history of the quiescent galaxy population is
unlikely to be simple, with different physical processes dominating at
different stellar masses and redshifts \citep[see
e.g. ][]{Choi:2014p9894,Brennan:2015p9469}. Given the number of free
parameters, studying the evolution of the distribution of galaxies in
star-formation-rate/stellar-mass/morphology planes can only help to
constrain the role of different processes in shutting down star
formation in combination with detailed models which help to link cause
and effect. A more direct method involves the identification of
galaxies caught in the act of transition: either those thought to be
in the final stage of a starburst
\citep{Nelson:2014p9939,Barro:2014p9893,vanDokkum:2015p10009}, or
galaxies that have recently quenched their star formation 
  \citep{Wild:2009p2609,Whitaker:2012p8738,Pattarakijwanich:2014p10261}. One
such class of transition galaxies are post-starburst galaxies, which
are identified by their (temporally resolved) recent star-formation
history. The presence of strong Balmer absorption lines, or a
significant Balmer break, in their spectra indicates an increased
fraction of A-F stars.  When these spectral features are strong
  enough, simulations show that this implies a recent, short and
  strong burst of star formation, followed by rapid truncation (Wild
  et al. 2009, hereafter WWJ09, Snyder et
  al. 2011)\nocite{Snyder:2011p10229}; weaker features may indicate
  truncation alone.

At low redshift, post-starburst galaxies have been linked to merger
events
\citep{Zabludoff:1996p9674,Blake:2004p9733,Goto:2005p9738,Yang:2008p9797,Pawlik:2016p10155},
and recently low-redshift post-starburst galaxies have been found to
have significant residual cold gas reservoirs
\citep{Zwaan:2013p9501,Rowlands:2015p9477,French:2015p9497}. Together
this implies that multiple merger events may be required to fully
quench star formation in low-redshift star-forming galaxies
\citep{Rowlands:2015p9477}. The formation of quiescent galaxies in the
present-day Universe may well follow a slower, less dramatic route
than at high redshift where higher gas fractions and plentiful
external gas supplies will lead to less stable disks, stronger
starbursts and AGN accretion, and stronger associated
outflows. Measuring the evolution in the number density, properties
and environment of post-starburst galaxies from the early Universe to
the present day may help us to understand the processes responsible
for shutting down star formation and building the red sequence.

Until recently, obtaining a large sample of post-starburst galaxies at
redshifts significantly greater than zero has been impossible, due to
the need for moderate signal-to-noise ratio rest-frame optical spectra
 \citep{2006ApJ...648..281Y,wild_psb}. Deep redshift surveys such
as the Vimos VLT Deep Survey (VVDS) and zCOSMOS have provided samples
of a few 10's at $z<1$ \citep[WWJ09,][]{Vergani:2010p9474}.  More
  recently, the Baryon Oscillation Spectroscopic Survey (BOSS) has
  provided a large number of spectroscopically selected post-starburst
  galaxies, but at the expense of a complex sample selection function
  \citep{Pattarakijwanich:2014p10261}.  In \citet[][hereafter
WA14]{Wild:2014p9476} we developed a new Principal Component Analysis
(PCA) based method to characterise the shape of the optical to
near-infrared (NIR) spectral energy distributions (SEDs) of
galaxies. Free from constraints of spectral synthesis models, this
allowed us to cleanly identify post-starburst galaxy candidates from
broad-band photometry alone, leading to samples of 100's at
$z>1$. Spectroscopic confirmation both through stacking (WA14) and
follow-up observations \citep{Maltby:2016p10156}, confirms the nature
of the photometrically selected post-starburst galaxies, with between
60 and 80\% spectroscopically confirmed as post-starburst galaxies,
depending on the specific criteria used.

The aim of this paper is to investigate the evolution in the number
density and mass function of photometrically-selected post-starburst
galaxies, from $z\sim2$ to $z\sim0.5$. Companion papers present
results on their morphology (Almaini et~al. submitted), spectroscopic
confirmation \citep{Maltby:2016p10156}, clustering (Wilkinson et~al. in
prep.), dust properties (Rowlands et~al. in prep.) and environment
(Socolovsky et~al. in prep.). 

Where necessary we assume a cosmology with $\Omega_M=0.3$,
$\Omega_\Lambda=0.7$ and $h=0.7$. All magnitudes are on the AB system
\citep{Oke:1983p10013}. Stellar masses are calculated assuming a
\citet{2003PASP..115..763C} initial mass function (IMF) and are
defined as the stellar mass remaining at the time of observation.

\section{Method} \label{sec:method}

We select galaxies brighter than a limiting $K$-band magnitude of 24,
from the UKIRT Infrared Deep Sky Survey (UKIDSS) Ultra Deep Survey
(UDS) Data Release 8 (DR8), a deep, large area near-infrared (NIR)
survey and the deepest of the UKIDSS surveys
\citep{Lawrence:2007p9799}. UKIRT observations provide $J$, $H$, $K$
observations to $5\sigma$ limiting depths in 2\arcsec\ diameter
apertures of 24.9, 24.2 and 24.6 mag respectively
\citep{Hartley:2013p9803}. Deep optical observations come from the
Subaru XMM-Newton Deep Survey \citep[SXDS, ][]{Furusawa:2008p8792}, to
depths of 27.2, 27.0, 27.0 and 26.0 in $V$, $R$, $i'$, $z'$. Mid-IR
coverage (IRAC 3.6\mum) is provided by the Spitzer UDS Legacy Program
(SpUDS, PI:Dunlop) to a depth of 24.2. Photometry was extracted within
3\arcsec\ diameter apertures at the position of the $K$-band sources,
with an aperture correction applied for the IRAC 3.6\mum\
image. Further details on the methods used to construct the UDS DR8
catalogue can be found in \citet{Hartley:2013p9803} and
\citet{Simpson:2012p9852}.

Photometric redshifts were computed following
\citet{Simpson:2013p9858}. Briefly, the multiwavelength SEDs were fit
with a linear combination of six solar metallicity, simple stellar
population templates with ages logarithmically spaced between 30\,Myr
and 10\,Gyr, and 3 templates dust-reddened using a Small Magellanic
Cloud extinction law.

After masking the survey area for bright stars and regions affected by
cross talk, the final science area is 0.62 square degrees. The UDS
field contains three large clusters at $z\sim0.6$ and $0.85$
\citep[SXDF69XGG, SXDF607XGG,
SXDF46XGG; ][]{vanBreukelen:2006p9860,Finoguenov:2010p9884}. As the
stellar population of galaxies in clusters differs from that in the
field, we tested the impact of removing galaxies potentially
associated with these clusters from our analysis, by removing the
areas of the survey in which these clusters lie. We found this had no
significant impact on our results, including on the evolution of the
fraction of quiescent galaxies.

From our $K<24$ catalogue, we select 48,713 galaxies with photometric
redshifts $0.5<z_{\rm phot}<2.0$. The lower redshift limit is applied
to ensure that the $V$-band provides a rest-frame flux measurement
bluewards of the 4000\AA/Balmer break. The slightly shallower depth of
the $B$ band in the UDS means that errors on determining the strength
and shape of the 4000\AA/Balmer break increase significantly below
this redshift. Similarly at high redshift, we find that we need a well
determined flux measurement redwards of rest-frame $\sim1$\mum\ to
accurately constrain the shape of the red end of the SED. At $z>2$,
the need for the Spitzer-IRAC channel 2 (4.5\micron) measurements caused
significantly larger errors on the determination of the shape of the
SEDs, and therefore on the estimation of the stellar populations of
the galaxies.

The UDS field also contains $\sim$3700 homogeneously observed spectra,
taken as part of the UDSz project using a combination of the VIMOS and
FORS2 instruments on the ESO VLT (ESO Large Programme 180.A-0776, PI:
Almaini). This allows for good determination of photometric redshifts,
as well as verification of the broad band SED analysis methods used in
this paper to determine the stellar population content of the
galaxies.  1,095 of the galaxies studied in this paper with
  $0.5<z<2.0$ have good quality spectroscopic redshifts from this
  programme, of which 207 are classified as quiescent or
  post-starburst according to their broad-band photometric SED shape
  (see next section).

\subsection{Super-colour analysis}\label{sec:SCmethod}

We follow the Principal Component Analysis (PCA) method of WA14 to
derive linear combinations of filters that describe the variation in
galaxy spectral energy distributions (SEDs) using only a small number
of components (eigenvectors). The eigenvectors are derived from a
  library of \citet[][hereafter BC03]{2003MNRAS.344.1000B} dust
  reddened spectral synthesis models, with ``stochastic''
  star-formation histories
  \citep{Kauffmann:2003p9798,Gallazzi:2005p6450}. From these models
  broad-band fluxes are calculated for each of the survey filters, in
  small steps of $\Delta z=0.01$ to cover the full redshift range of
  the data.  A PCA of these super-sampled SEDs is performed, which
  links together observed bands that vary together to compress the
  data into a small number of eigenvectors. We find that only 3
  eigenvectors are required to account for $>99.9$\% of the variance
  in our library of model SEDs. The eigenvectors are weights, which
  are then used to sum the observed fluxes of a galaxy, in an optimal
  way to highlight the differences in shapes of galaxy SEDs. For each
  individual observed galaxy, only a small number of points in the
  weighting arrays are used, appropriate for the particular redshift
  of the galaxy. The sparseness of the observed data leads to larger,
  but well defined, errors on the resulting principal component
  amplitudes \citep{1999AJ....117.2052C}.

 The amplitude of each of the eigenvectors within a galaxy's SED
 (termed ``super-colour'') is exactly equivalent to a traditional
 ``colour'', in the sense that it is a linear combination of flux
 measured in different filters. However, by combining all the
 available photometric bands into this ``super-colour'' in an optimal
 way through a weighting vector, the full information contained within
 the multi-wavelength surveys is retained. Equally, the observations
 are not forced to match spectral synthesis models, as is the case when SED
 fitting is used to estimate rest-frame colours. This freedom from the
 spectral synthesis models allows us to easily identify regions of SED
 space that are not well described by the models, and where physical
 parameters derived from model fitting will be biased.

 The method does not solve for the redshift of the galaxy, but takes
 as input the redshift from a standard photometric redshift code
 \citep{Simpson:2013p9858}. This was a practical decision taken
 because current photometric redshift codes are highly optimised for
 the task, and repeated effort seemed unnecessary. The effect of
 reasonable photometric redshift errors on the observed super-colours
 is negligible (see Figure 5 of WA14). We have also verified, using
 the spectroscopic redshifts from the UDSz project and from Maltby
 et~al. (2016), that there are no systematic biases caused by
 photometric redshift errors in each of the SED classes and redshift
 bins studied in this paper. The redshift errors of both the quiescent
 and post-starburst populations are low ($\sigma_z/(1+z)=0.02$) and
 are constant with both spectral type and redshift over the redshift
 range studied here.

\begin{figure}
  \includegraphics[scale=0.4]{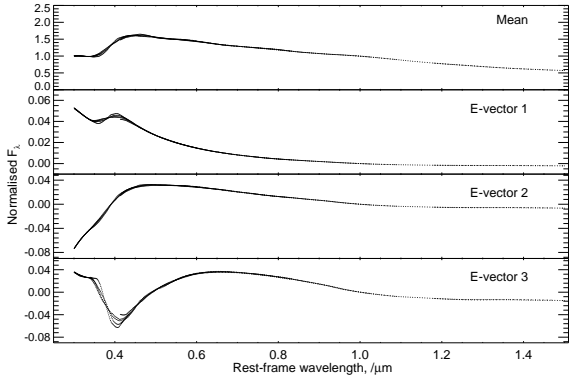}
  \caption{The mean and first 3 eigenvectors from a Principal
    Component Analysis of a library of Bruzual \& Charlot (2003)
    ``stochastic burst'' model SEDs, convolved with broad band filter
    functions to match the UDS survey observations for galaxies in the
    redshift range $0.5<z<2.0$, in steps of $\Delta z = 0.01$. Each
    point represents the weight given to a particular filter for a
    galaxy at a particular redshift. The points are placed at the
    central rest-frame wavelength of the filter and the y-axis
    indicates rest-frame flux. The multiple curves reflect the
    fact that different filters may probe the same rest-frame wavelength of
    galaxies at different redshifts. On addition to the
    mean, the first eigenvector primarily alters the red-blue slope of
    the SED, the second and third primarily alter the shape of the
    4000\AA/Balmer break region.} \label{fig:evecs}
\end{figure}

 The full method is described in detail in WA14, the only change made
 in this paper is to extend the redshift range to $0.5<z<2.0$. This
 change results in quantitatively different super-colour values, but
 no qualitative difference in results.  The new eigenvectors are
 presented in Figure \ref{fig:evecs} and can be seen to differ only
 slightly from those in Figure 2 of WA14. Each point represents the
 weight given to one filter in our filter set, used to observe a
 galaxy at the given rest-frame wavelength. The multiple curves
 reflect the fact that different filters (with correspondingly
 different transmission functions) will be used to observe the same
 rest-frame wavelength of galaxies at different redshifts. For the
 purposes of the analysis in this paper, the galaxies have been
 separated into three classes based on the shapes of their SEDs
 defined by the first three super-colours: quiescent, star-forming and
 post-starburst (PSB). Compared to WA14, this combines all the
 star-forming classes into a single class, and adds the candidate
 low-metallicity objects into the quiescent class.

\subsection{Stellar masses and volume correction}\label{sec:mass}
Stellar mass-to-light ratios were calculated for each galaxy using a
Bayesian analysis which accounts for the degeneracy between physical
parameters. Specifically, we fit a library of 10's of
thousands\footnote{Between 23,000 and 62,000 depending on the redshift
  of the galaxy.} of \citet{2003MNRAS.344.1000B} population synthesis
models with a wide range of star-formation histories, two-component
dust contents and metallicities to the first three super-colours, to
obtain a probability density function (PDF) for each physical
property. We extracted the median, 16th and 84th of the PDF to
estimate the mass-to-light ratio and corresponding errors. The assumed
model star-formation histories are exponentially declining with
superimposed random starbursts with priors as described in
\citet{Kauffmann:2003p9798} and \cite{Gallazzi:2005p6450}. Resulting
stellar masses were increased by 10\% to allow for light lost from the
3\arcsec\ apertures used to extract the photometry. 

Determining the exact stellar masses of galaxies is difficult, and
differences between methods typically lead to 0.3\,dex systematic
uncertainties. We verified that all the results presented here are
robust to different methods for obtaining stellar mass estimates by
rerunning our analysis using masses from both
\citet{Simpson:2013p9858} and \citet{Hartley:2013p9803}. However, all
of these analyses rely on \citet{2003MNRAS.344.1000B} population
synthesis models. For quiescent galaxies at high-redshift, additional
errors on stellar masses may arise from the less well constrained
contribution of TP-AGB stars to the NIR portion of the stellar
continuum \citep{Maraston:2006p10268,Conroy:2009p9193}. For our
results, it is important that there are no systematic biases in masses
between populations and with redshift. The former could arise because
the contribution of TP-AGB stars changes as a function of stellar
population age. \citet{Maraston:2006p10268} found that quiescent
galaxies with BC03 ages $\gsim$1\,Gyr have younger ages,
and therefore lower mass-to-light ratios, when fit with models with a
significantly greater contribution from the TP-AGB phase.  A
systematic error on the stellar masses with redshift could also arise
as different bands are used to probe the red end of the SED; in
particular the use of the IRAC 3.6\mum\ band for galaxies at $z>1.4$
could lead to a bias, due to the lower spatial resolution of these
observations. We will note the impact of these uncertainties on our
results where relevant.

We also used the best-fit (minimum $\chi^2$) population synthesis
model, together with the SED normalisation estimated during the
super-colour analysis, to determine the fraction of the total survey
volume in which a galaxy of that SED shape and absolute magnitude
would be observed, given the observed frame $K$-band flux limit of the
survey. Stellar mass functions were corrected for these volume effects
using the V$_{\rm max}$ method which weights each galaxy by 1/V$_{\rm
  max}$, where V$_{\rm max}$ is the maximum volume in which the galaxy
may be observed in the survey \citep{1968ApJ...151..393S}. Mass
completeness limits in each redshift slice and for each spectral type
were determined following the method of \citet{Pozzetti:2010p9506},
allowing for a 90\% completeness factor.

\section{Results}\label{sec:results}

\begin{figure*}
  \includegraphics[scale=0.9]{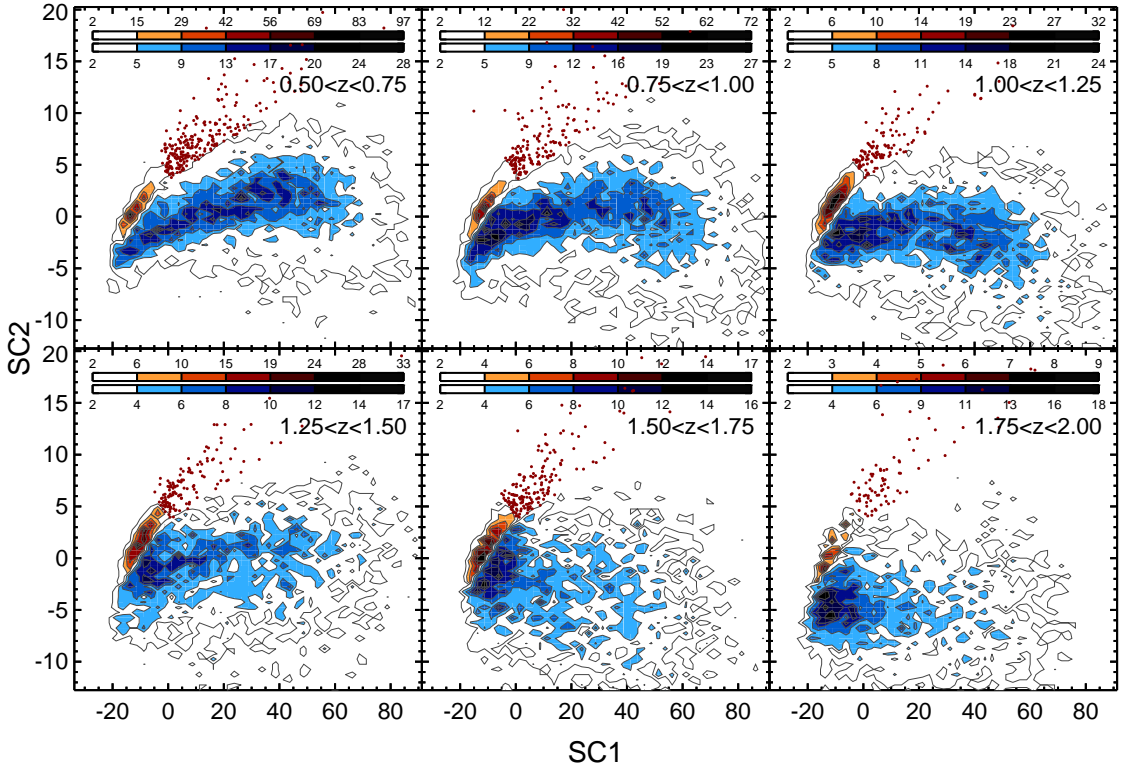}
  \caption{The distribution of the first two super-colours (SC) for
    galaxies observed in the UDS field, in six  photometric-redshift intervals as
    indicated in the top-right of each panel. The filled contours and
    colour bars at the top indicate the number of galaxies in bins of
    $\Delta$SC1$\times\Delta$SC2 of $2.5\times0.5$,  individually for
    the quiescent and star-forming populations. Note that the
    plots (and associated numbers) are as observed and have not been
    volume corrected.  The tight red sequence (red contours) can be
    observed to the left of SC1/2, with the blue-cloud (blue contours)
    extending from low to high values of SC1.  The PSB
    galaxies (filled brown circles) are seen as a well-defined stream
    entering the top end of the red sequence. The third super-colour
    (not shown here) has been used to exclude a small number of
    candidate low-metallicity quiescent galaxies, which can otherwise
    be interlopers in PSB samples. } \label{fig:SC12_redshift}
\end{figure*}

In Fig. \ref{fig:SC12_redshift} we present the colour-colour
distributions for the first two super-colours in six redshift
intervals. Based on both comparison to spectral synthesis models, and
spectroscopic confirmation, we showed in WA14 how different types of
galaxies lie in different regions of super-colour space. Galaxies
which have undergone no significant recent star formation (quiescent
galaxies) form a tight and well defined ``red sequence'' with low
SC1. The star-forming population of galaxies covers a wide range of
SC1, from the very bluest objects with SC1$\sim70$ to a small number
of galaxies that are redder than quiescent galaxies. As shown in WA14
by stacking rest-frame optical spectra, the extremely red galaxies
that fall below the red sequence in SC2 (which correlates with
4000\AA\ or Balmer break strength) are truly star-forming with stellar
absorption lines and nebular emission lines consistent with a young
stellar population. We showed in WA14 that their red optical-to-NIR SEDs
are consistent with high dust contents and metallicities, while their
mean stellar ages are similar to ``typical'' star-forming galaxies at
that redshift. Note that these plots are based on a $K$-band limited
sample and are not volume corrected, and therefore a different mass
distribution of galaxies is visible at different redshifts. This is
most obvious in the blue sequence, where we only see the most massive
blue sequence galaxies in the highest redshift bins. Their old ages,
high metallicities and/or high dust contents make them appear towards the
red end of the blue sequence.

According to our comparison to spectral synthesis models, and
spectroscopic confirmation in both WA14 and \citet{Maltby:2016p10156},
PSB galaxies form a well-defined stream entering the top of the red
sequence, clearly separated above the blue sequence by their
distinctive strong Balmer break (i.e. high SC2 values). The exact
boundaries between the PSB and quiescent/star-forming galaxies are
somewhat arbitrary; here we define them based on our ability to
identify PSB galaxies from their strong Balmer absorption lines in
moderate resolution rest-frame optical spectra (WA14). Varying the
boundaries slightly does not have any significant impact on our
results.

\subsection{Decline in fraction and number density of quiescent,
  star-forming and PSB galaxies}\label{sec:numberdens}

\begin{figure*}
\hspace*{-0.5cm} 
\includegraphics[scale=0.55]{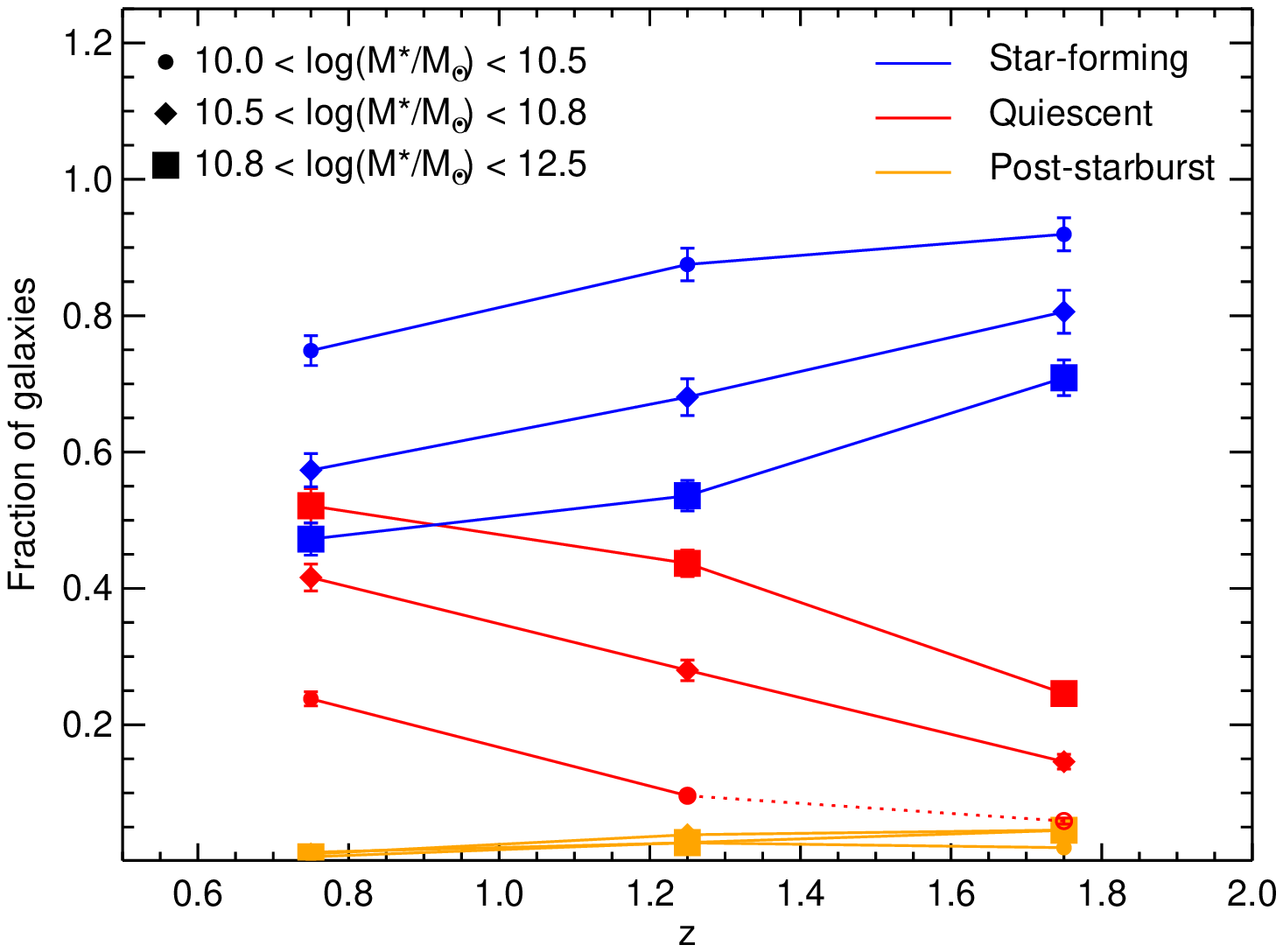}
  \includegraphics[scale=0.55]{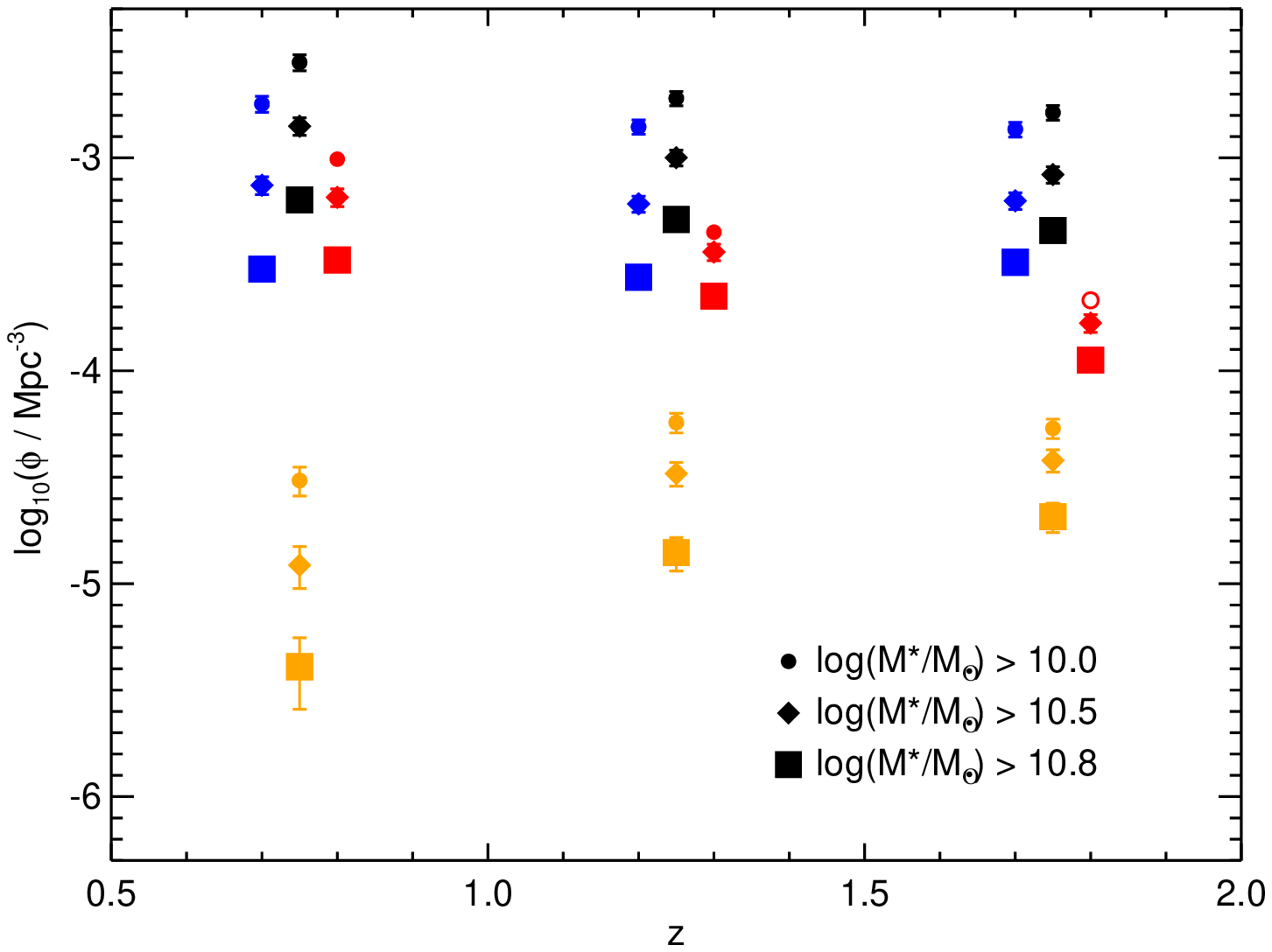}
  \caption{{\it Left:} The  volume corrected fraction of quiescent (red), star-forming
    (blue) and PSB (orange) galaxies as a function of redshift and
    stellar mass, as indicated by the key in the top-left. Errors are
    propagated from the Poisson errors in each bin in the usual
    way. {\it Right:} The number density of all (black), quiescent
    (red), star-forming (blue) and PSB (orange) galaxies as a function
    of redshift and mass. The mass limits are given in the
    bottom-right of the figure.  Errors include contributions from
    Poisson errors and cosmic variance.  The blue and red points have
    been offset slightly on the x-axis for clarity. Note that the
    high-redshift quiescent sample has a 90\% mass completeness limit
    of \logM$=10.2$, and therefore is plotted as an open
    symbol. }  \label{fig:FractionNumber_z}
\end{figure*}

\begin{table}
\begin{center}
  \caption{The comoving log number density (Mpc$^{-3}$) for all, star-forming,
    quiescent and PSB galaxies in redshift and stellar mass
    bins. The 90\% completeness limits for each bin are given in Table \ref{tab:schechter}. Note that the 
    high-redshift quiescent sample has a 90\% mass completeness limit of
    \logM$=10.2$, and
    therefore the number density is given in brackets. Errors include
    both Poisson and cosmic variance contributions.
  }\label{tab:number}
\begin{tabular}{cccc}\hline\hline
 & 0.5$>z>$1.0&1.0$>z>$1.5&1.5$>z>$2.0\\ \hline
\multicolumn{4}{c}{All} \\
\hline
$\log({\rm M}/{\rm M}_\odot) >$ 10.0
 & $-2.6\pm 0.04$
 & $-2.7\pm 0.03$
 & $-2.8\pm 0.03$
\\
$\log({\rm M}/{\rm M}_\odot) >$ 10.5
 & $-2.9\pm 0.04$
 & $-3.0\pm 0.04$
 & $-3.1\pm 0.04$
\\
$\log({\rm M}/{\rm M}_\odot) >$ 10.8
 & $-3.2\pm 0.05$
 & $-3.3\pm 0.04$
 & $-3.3\pm 0.05$
\\
\hline
\multicolumn{4}{c}{Quiescent} \\
\hline
$\log({\rm M}/{\rm M}_\odot) >$ 10.0
 & $-3.0\pm 0.04$
 & $-3.3\pm 0.03$
 & $ (-3.7\pm 0.04) $
\\
$\log({\rm M}/{\rm M}_\odot) >$ 10.5
 & $-3.2\pm 0.04$
 & $-3.4\pm 0.04$
 & $-3.8\pm 0.04$
\\
$\log({\rm M}/{\rm M}_\odot) >$ 10.8
 & $-3.5\pm 0.05$
 & $-3.6\pm 0.05$
 & $-3.9\pm 0.05$
\\
\hline
\multicolumn{4}{c}{Star-forming} \\
\hline
$\log({\rm M}/{\rm M}_\odot) >$ 10.0
 & $-2.7\pm 0.04$
 & $-2.9\pm 0.03$
 & $-2.9\pm 0.03$
\\
$\log({\rm M}/{\rm M}_\odot) >$ 10.5
 & $-3.1\pm 0.04$
 & $-3.2\pm 0.04$
 & $-3.2\pm 0.04$
\\
$\log({\rm M}/{\rm M}_\odot) >$ 10.8
 & $-3.5\pm 0.05$
 & $-3.6\pm 0.05$
 & $-3.5\pm 0.05$
\\
\hline
\multicolumn{4}{c}{Post-starburst} \\
\hline
$\log({\rm M}/{\rm M}_\odot) >$ 10.0
 & $-4.5\pm 0.07$
 & $-4.2\pm 0.05$
 & $-4.3\pm 0.05$
\\
$\log({\rm M}/{\rm M}_\odot) >$ 10.5
 & $-4.9\pm 0.10$
 & $-4.5\pm 0.06$
 & $-4.4\pm 0.05$
\\
$\log({\rm M}/{\rm M}_\odot) >$ 10.8
 & $-5.4\pm 0.16$
 & $-4.9\pm 0.08$
 & $-4.7\pm 0.07$
\\

\hline
\end{tabular}
\end{center}
\end{table}

In the left panel of Fig. \ref{fig:FractionNumber_z} we present the
fraction of quiescent, star-forming and PSB galaxies as a function of
redshift and stellar mass. We see the now familiar evolution in the
bimodality in the galaxy population, with an increase in quiescent
galaxy fraction with decreasing redshift at all stellar masses. This
figure also shows how rare PSB galaxies are at all redshifts: the
fraction of PSB galaxies declines from around 5\% of the total
population in any given mass bin at $z\sim2$, to 1\% by $z\sim0.5$. It
is important to remember that while the red fraction is cumulative,
(i.e. galaxies are expected to move onto the red sequence and stay
there, at least on average), the PSB phase is a transitory phase
through which the galaxies move on timescales of several hundred Myr
to a Gyr. Therefore, the scarcity of PSB galaxies relative to
quiescent and star-forming galaxies does not automatically imply that
they are an unimportant phase of galaxy evolution.

The right panel of Fig. \ref{fig:FractionNumber_z} shows the number
density of all, quiescent, star-forming and PSB galaxies as a function
of redshift with masses above different limits. The values are
provided in Table \ref{tab:number}.  Errors are calculated by
combining the Poisson errors together with a contribution from cosmic
variance, following the prescription of
\citet{Moster:2011p10145}. Errors caused by photometric errors leading
to different  photometric redshifts and super-colour
  classifications were found to be insignificant and are not
included. We see the now familiar result, that the number density of
quiescent galaxies grows with time, at all stellar masses, while the
number density of the star-forming galaxies remains almost constant
\citep[e.g.][]{2007ApJ...665..265F,Ilbert:2013p9113,
  Muzzin:2013p9592}. This increase in the number density of massive
quiescent galaxies implies the transition of galaxies from the blue to
the red sequence. On the other hand, the number density of galaxies
observed in a PSB phase decreases with time, particularly in the
higher mass limit bins, and at a redshift below $z\lsim1$.

The number density of PSB galaxies identified spectroscopically from
the VVDS survey with $0.5<z<1.0$ by WWJ09 was \logphi$=-4.0$ for
\logM$> 9.75$, measured from 18 observed PSB galaxies. For comparison,
we compute the number density using the same mass limit and redshift
range and find \logphi$=-4.35$, measured from 86 galaxies which is
significantly lower assuming Poisson errors alone\footnote{In
  the z-COSMOS field \citet{Vergani:2010p9474} found a number density
  of \logphi$=-4.3$ for spectroscopically identified PSB galaxies,
  using a similar redshift range to WWJ09 but with a slightly higher
  mass cut \logM$ > 10$ and different spectroscopic selection
  criteria. For this mass cut, we again find a number density that is
  a little lower than when using a spectroscopic selection at
  \logphi$=-4.5$.}.  The fact that we identify fewer PSB galaxies using
the photometric selection is unsurprising as we expect photometric
selection to be less sensitive to the PSB spectral features than
spectroscopic observations i.e. the photometric method should detect
the PSBs with the strongest features, which will be the youngest
and/or those with the largest burst-to-old stellar mass ratio.
The overall completeness of the super-colour method for detecting PSB
galaxies compared to spectroscopic searches is difficult to ascertain
without large, blind spectroscopic surveys for comparison. However, it
is estimated by \citet{Maltby:2016p10156} to be $\sim$60\%, in good
agreement with the difference in number densities found here. We note
that this interpretation disagrees with the much lower number
densities of post-starburst galaxies detected in the spectroscopic
BOSS survey by \citet{Pattarakijwanich:2014p10261}, where the
luminosity function lies about 1\,dex below ours as presented in
WA14. Understanding the effect of photometric vs. spectroscopic
selection of post-starburst galaxies on the resulting star-formation
histories is a topic of ongoing investigation. The potential for
significant differences should be kept in mind when comparing our
results with those from current and future spectroscopic surveys.

\subsection{Mass function evolution of quiescent, star-forming and PSB
galaxies}\label{sec:mf}

\begin{figure*}
  \includegraphics[scale=0.7]{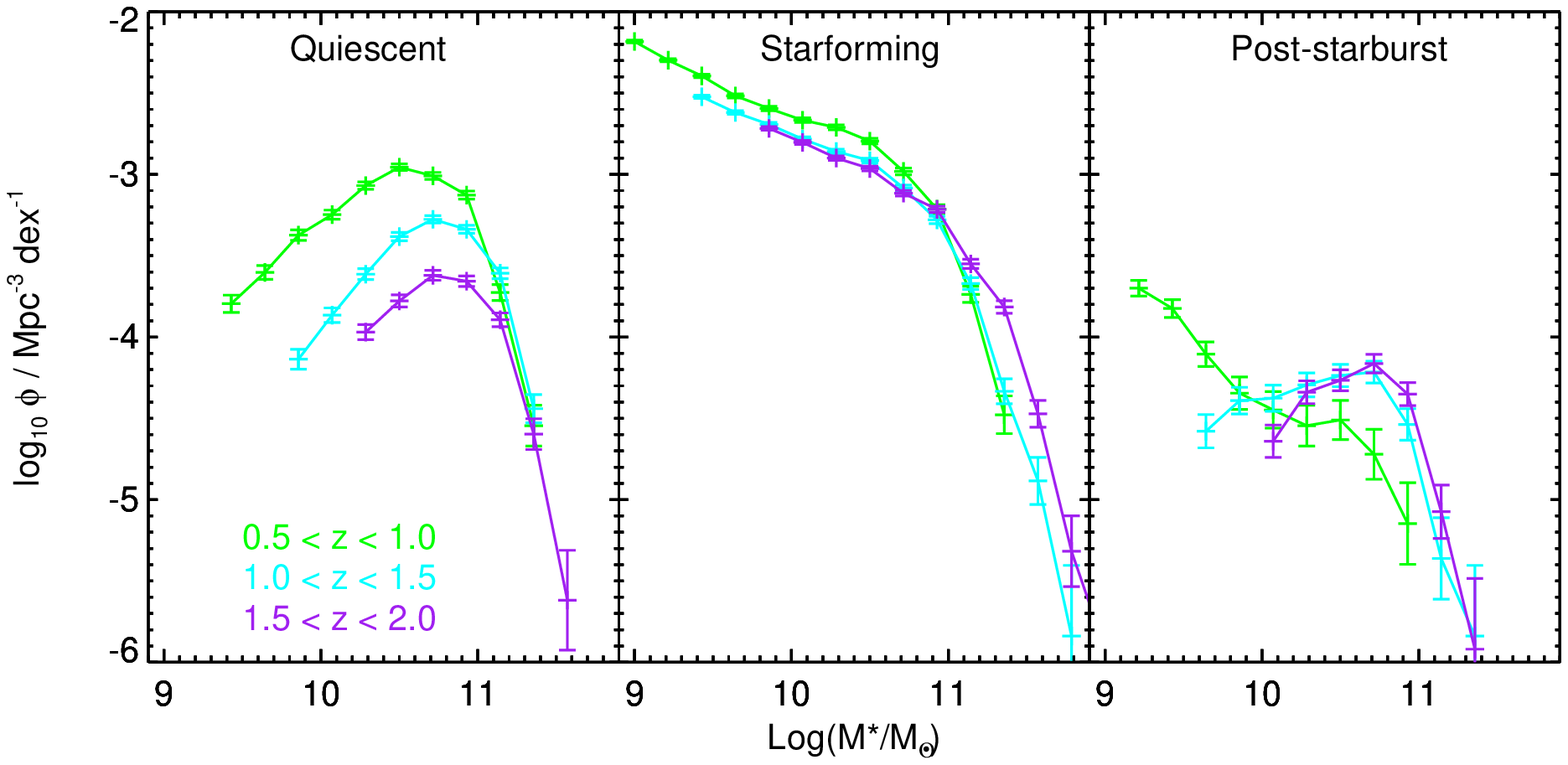}\\
  \includegraphics[scale=0.7]{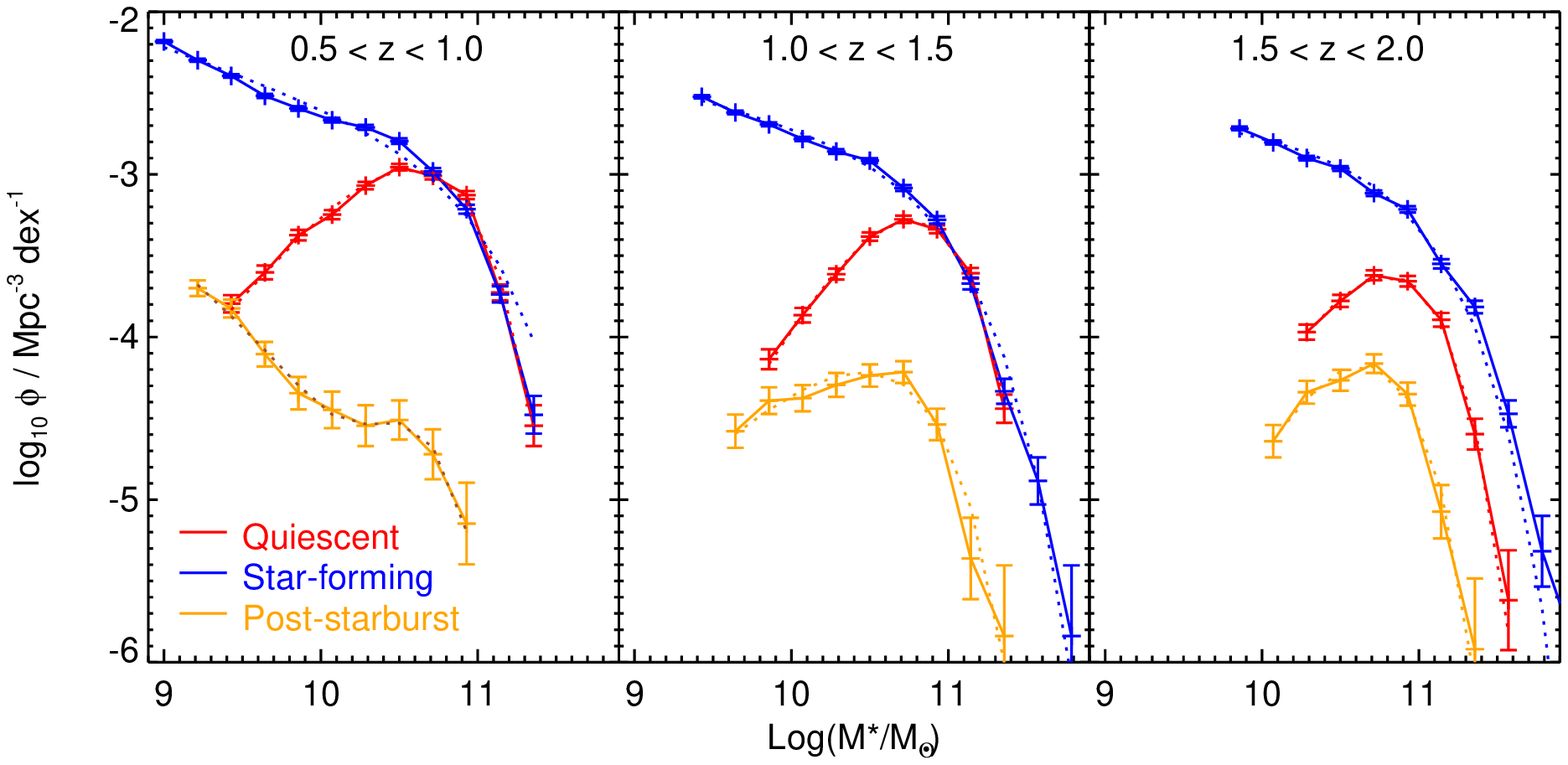}\\
  \includegraphics[scale=0.7]{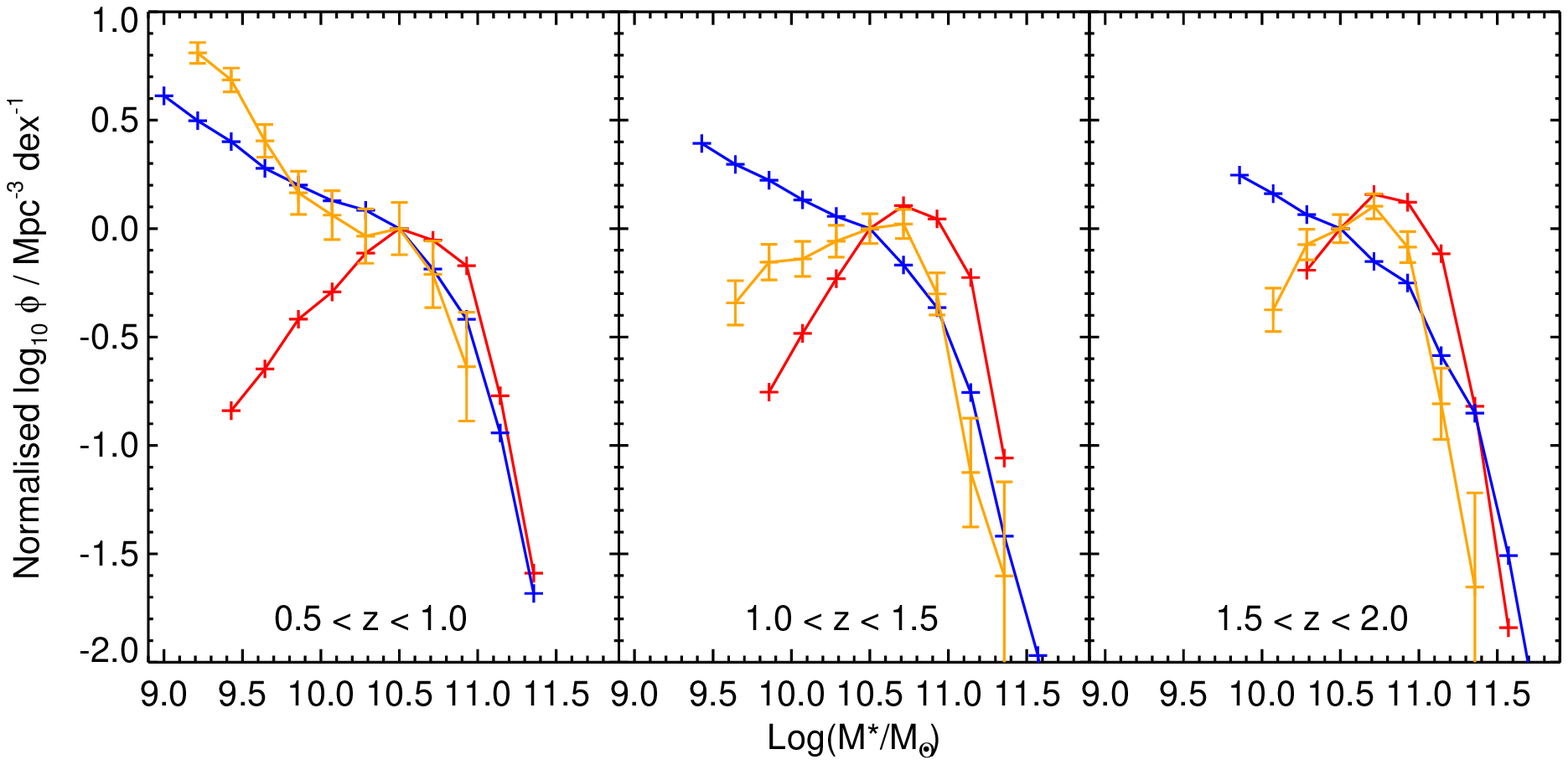}
  \caption{{\it Top and centre:} The stellar mass functions of
    quiescent, star-forming and PSB galaxies as a function
    of redshift.  In the central panel, the overplotted dotted lines
    are the Schechter function fits. {\it Bottom:} the mass functions have been
    normalised at \logM$=10.5$ in order to better compare their
    shape. Errors include both Poisson and cosmic variance contributions.} \label{fig:MF}
\end{figure*}

\begin{table}
\begin{center}
  \caption{The Schechter function parameters fitted to the quiescent,
    star-forming and PSB populations. The first line in
    each section gives the 
    90\% mass completeness limits in each bin, in \logM. Only for the lowest
    redshift bin of the PSB population was it necessary to
    fit a double Schechter function. 
  }\label{tab:schechter}
\begin{tabular}{cccc}\hline\hline
 & 0.5$<z<$1.0&1.0$<z<$1.5&1.5$<z<$2.0\\ \hline
\multicolumn{4}{c}{Quiescent} \\
\hline
Compl. Lim.
&9.38
&9.78
&10.17
\\
$\log({\rm M^*}/{\rm M}_\odot)$
& $10.52\pm$0.02
& $10.55\pm$0.02
& $10.59\pm$0.03
\\
$\log(\phi^*/{\rm Mpc}^{-3})$
& $-2.90\pm$0.01
& $-3.28\pm$0.01
& $-3.62\pm$0.03
\\
$\alpha$
& $0.16\pm$0.04
& $0.68\pm$0.08
& $0.65\pm$0.15
\\
\hline
\multicolumn{4}{c}{Star-forming} \\
\hline
Compl. Lim.
&8.92
&9.29
&9.80
\\
$\log({\rm M^*}/{\rm M}_\odot)$
&11.00$\pm$0.00
&10.95$\pm$0.02
&11.00$\pm$0.00
\\
$\log(\phi^*/{\rm Mpc}^{-3})$
& $-3.27\pm$0.01
& $-3.27\pm$0.02
& $-3.26\pm$0.01
\\
$\alpha$
& $-1.35\pm$0.01
& $-1.25\pm$0.02
& $-1.16\pm$0.01
\\
\hline
\multicolumn{4}{c}{Post starburst} \\
\hline
Compl. Lim.
&9.10
&9.56
&9.88
\\
$\log({\rm M^*}/{\rm M}_\odot)$
&10.15$\pm$0.20
&10.53$\pm$0.07
&10.39$\pm$0.06
\\
$\log(\phi^*_1/{\rm Mpc}^{-3})$
& $-4.79\pm$0.43
& $-4.15\pm$0.05
& $-4.21\pm$0.05
\\
$\alpha_1$
& $1.37\pm$1.29
& $-0.15\pm$0.15
& $0.79\pm$0.27
\\
$\log(\phi^*_2/{\rm Mpc}^{-3})$
& $-4.70\pm$0.36& -- & -- 
\\
$\alpha_2$
& $-1.76\pm$0.25& -- & -- 
\\
\hline

\hline
\end{tabular}
\end{center}
\end{table}

Further information on the evolution of the populations can be
obtained from the galaxy stellar mass functions.  In Fig.
\ref{fig:MF} we present the stellar mass functions of the
star-forming, quiescent and PSB populations as a function of redshift
(top and middle panels). In the lower panel we compare the shape of
the mass-functions by normalising them at \logM$=10.5$. Errors include
both Poisson and cosmic variance contributions.  We fit single
  Schechter functions \citep{Schechter:1976p9956} to all mass
  functions, apart from the lowest redshift PSB bin, for which the
  upturn at lower stellar masses required a double Schechter function
  \citep{Pozzetti:2010p9506}. The results are given in Table
  \ref{tab:schechter} and all the trends noted below are robust to
  binning. No allowance is made for Eddington bias which can scatter
  galaxies into the massive end of the mass function, as our results
  do not require accurate number densities for the highest mass
  objects. \citet{Caputi:2011p10153} suggest that by not accounting
  for Eddington bias the number densities of the highest mass galaxies
  [\logM$>11$] will be over estimated by 0.13\,dex.  If there were a
  significantly greater contribution from the TP-AGB phase of stellar
  evolution to the SEDs of our galaxies (Section \ref{sec:mass}), this
  would act to move the quiescent mass functions to lower masses,
  relative to the other spectral classes. In
  our highest redshift bin, the masses rely significantly on the IRAC
  3.6\mum\ band which has larger uncertainties than the other
  bands. Using smaller redshift bin sizes we confirmed that our mass
  functions do not show any significant change in shape as this band
  is introduced. 

The quiescent and star-forming stellar mass functions show the now
familiar build up in the number density of quiescent galaxies,
particularly at intermediate and low masses, but little change in the
number density of star-forming galaxies at any mass. The star-forming
stellar mass function has a relatively constant characteristic mass
($M*$), number density ($\phi*$) and faint end slope parameter
($\alpha$), that are consistent with previous measurements
\citep{Muzzin:2013p9592}. The quiescent galaxy stellar mass function
turns over at intermediate masses, resulting in a positive value for
the faint end slope parameter. While the number density of quiescent
galaxies increases with cosmic time, the faint end slope flattens.
Unfortunately, the measured value of $\alpha$ for the quiescent
galaxies varies significantly between surveys and analysis methods,
making it difficult to make direct comparisons. However, the general
picture of a flattening with time is robust, with a tentative upturn
appearing at lower stellar masses
\citep{Drory:2009p10269,Muzzin:2013p9592,Tomczak:2014p10010}.

Interestingly, the shape of the PSB mass function is similar to that
of quiescent galaxies at high redshift, but the low mass end builds up
rapidly with cosmic time, leading the mass function to be more similar
to that of the star-forming galaxies at $z\lsim1$. This trend clearly
continues to lower redshift, as seen in the the luminosity function of
local post-starburst galaxies \citep{2004ApJ...602..190Q,
  Pattarakijwanich:2014p10261}. The turn over in the mass functions at
$z>1$ suggests that the PSB phenomenon is reserved for high-mass
galaxies alone at these redshifts, whereas at $z<1$ there is a
complete reversal with PSB spectral features predominantly being found
in low mass galaxies. As we will discuss in the following section, the
changing shape of the mass function with redshift suggests there may
be two different evolutionary pathways to form PSB galaxies.

\section{Discussion}

There are many physical processes that have been proposed to play a
role in the evolution of the demographics of the galaxy population
over cosmic time. Both ``fast'' ($\sim10^8$\,yr) and ``slow''
($\gsim10^9$\,yr) quenching routes may exist to convert star-forming
galaxies into quiescent galaxies, and thus build the bimodality in
galaxy populations observed in the present day Universe. For the
``fast'' quenching route, both ``starburst'' and ``switch-off'' modes
may play a role, with the former implying that a significant fraction
of stars are formed in a starburst that is subsequently truncated
along with star formation in the whole of the galaxy, and the latter
implying that no starburst occurs and the galaxy's star formation is
simply rapidly truncated. It is certainly possible, if not likely,
that the balance of all of these mechanisms varies with cosmic time.

In this paper we focus on the ``fast-quenching'' pathway, as this is
the route that our detection method is sensitive to. The A/F-star
features that allow us to identify the quenched galaxies are only
visible if there is a significant deficit in O/B stars, which requires
quenching timescales significantly shorter than the main sequence
lifetime of A/F stars. We begin this discussion by asking whether our
results are consistent with all quiescent galaxies at $z>0.5$ forming
following a fast-quenching episode. We then compare with spectral
synthesis models to further investigate the likely star formation
histories of the PSB class of galaxies. Based on the stellar mass
functions of the PSB galaxies, we propose that two different formation
mechnisms dominate at high and low redshift. Finally, we discuss more
speculatively whether high-redshift PSB galaxies could be the
descendants of sub-mm galaxies.

\subsection{The contribution of PSBs to the build up of the red
  sequence at \boldmath{\logM$>10$ and $z<2$}}

\begin{figure}
  \includegraphics[scale=0.55]{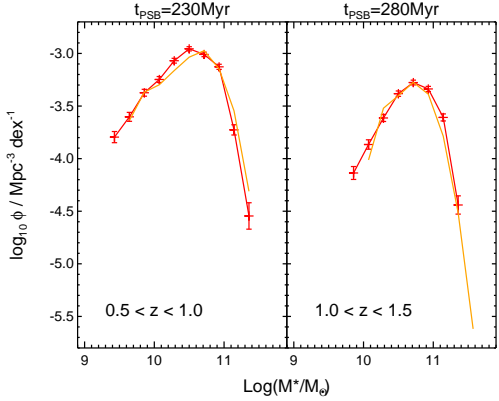}
  \caption{As the PSB galaxies are expected to enter the red sequence
    on timescales shorter than the difference in cosmic time between
    our redshift bins, we can combine the PSB and quiescent mass
    functions in a given redshift bin to predict the quiescent mass
    function in the next lowest redshift bin. {\it Red lines:} The
    stellar mass functions of quiescent galaxies in our lowest two
    redshift bins. {\it Orange lines:} The stellar mass functions of
    quiescent galaxies predicted from adding the stellar mass function
    of quiescent galaxies in the next highest redshift bin to the mass
    function of PSB galaxies in that redshift bin, multiplied by a
    factor which is fitted for. This factor can be translated directly
    into a minimum PSB visibility timescale, given in the title of
    each panel, after allowing for the difference in cosmic time
    between the redshift bins. } \label{fig:PSBpredict}
\end{figure}

\begin{figure}
  \includegraphics[scale=0.55]{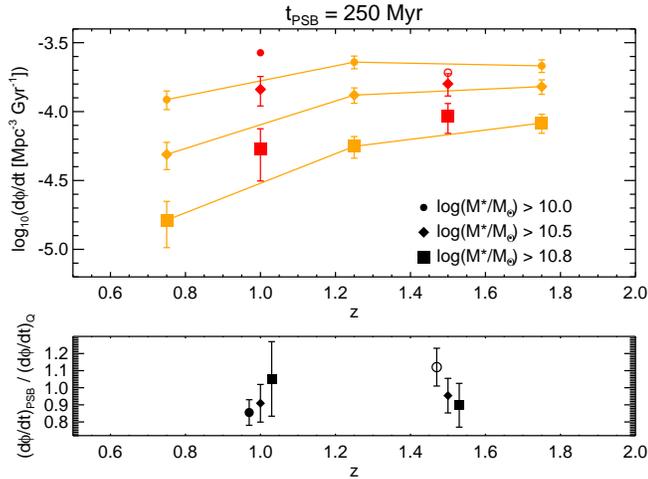}
  \caption{ {\it Top:} The number growth rate of quiescent galaxies
    (red) for the mass limits as given in the key, measured from the
    gradient in their number density evolution.  This is compared to
    the rate at which galaxies pass through a PSB phase (orange),
    assuming that this phase is observable for 250\,Myr (see text). Errors include
    contributions from both Poisson errors and cosmic variance. {\it
      Bottom:} The fraction of the number growth of quiescent galaxies
    that could be contributed by galaxies passing through a PSB
    phase. Poisson errors only are
    propagated in the standard way. The different measurements are
    artificially offset from one another on the x-axis for
    clarity.} \label{fig:number_z}
\end{figure}

Based on our understanding of the timescale during which A/F star
features are visible following a rapid truncation in star formation,
post-starburst galaxies are expected to join the quiescent population
within $\sim$1\,Gyr, which is similar to the time between our three
redshift bins. Therefore, we can add the PSB mass function to the
quiescent mass function in the same redshift bin, fitting for a
constant multiplicative factor that accounts for the difference in
cosmic time between the bins and the visibility
timescale of the PSB features, and
compare the resulting mass function with the subsequent redshift
bin. This allows us to (1) constrain the minimum visibility timescale
for the spectral features, and (2) investigate whether the mass
function of any remaining ``slow-quenched'' population can be
different to the PSB mass function. The result is shown in Figure
\ref{fig:PSBpredict}. If 100\% of the growth of the quiescent
population arises following a detectable PSB phase, we find that the PSB
features are visible for $\sim$250\,Myr.  This is entirely consistent
with the minimum timescale derived by WWJ09 through comparison to
hydrodynamic simulations of major gas-rich galaxy mergers. Clearly,
the shapes of the predicted mass functions in both redshift bins are
also in very good agreement with the actual quiescent galaxy mass
functions, which is required if 100\% of the growth of the quiescent
population does indeed arise following rapid truncation of the star
formation.

Our results show that any additional ``slow-quenching'' mechanism that
exists and contributes significantly to the mass growth of the
quiescent population with \logM$>10$ at $z>0.5$, must have the same
efficiency as a function of mass as the fast-quenching mechanism, in
order to ensure that the final quiescent mass function matches that
observed. If the ``slow-quenching'' route contributes significantly to
the build up of the quiescent population, the differing shapes of the
star-forming and quiescent mass functions imply that star-forming
galaxies must use up their gas supplies and halt their star formation
at a rate that is dependent on their stellar mass at $z>1$, to avoid
an excessive build up of the low mass end of the quiescent mass
function.

The same result can be depicted slightly differently, by using the
gradient in the number density evolution of the quiescent galaxies
(Fig. \ref{fig:FractionNumber_z}), to calculate the rate at which new
galaxies enter the red sequence ($d\phi/dt$). This is shown in the top
panel of Fig.  \ref{fig:number_z} (red symbols) for three different
mass cuts, and is compared to the rate at which galaxies are passing
through a PSB phase (orange symbols), assuming a visibility timescale
of 250\,Myr as estimated above.

We find that a few times $10^{-4}$ galaxies per Mpc$^3$ with
\logM$>10$ enter the red sequence every Gyr, although the precise
number depends on stellar mass and redshift. For \logM$>10.8$, the
rate at which galaxies are entering the red sequence is decreasing
with cosmic time from 9.2 to $5.3\times10^{-5}$ galaxies per Mpc$^3$
per Gyr between $z\sim1.5$ and $z\sim1$. For \logM$>10$, the rate at
which galaxies are entering the red sequence appears to be increasing
slightly with cosmic time, from 1.9 to $2.6\times10^{-4}$ galaxies per
Mpc$^3$ per Gyr. It is possible that this evolution is consistent with
being flat, however, given the slightly higher incompleteness in our
highest redshift, lowest mass bin. This mass trend is one
manifestation of the downsizing observed in the galaxy population
\citep[see][for a review of the literature]{Fontanot:2009p10005}, with
higher mass quiescent galaxies being formed predominantly at higher
redshifts, while galaxies continue to enter the low-mass end of the
quiescent population to the present day.

In the bottom panel of Fig. \ref{fig:number_z} we divide the number
growth rate ($d\phi/dt$) for the PSB galaxies (using a linear
interpolation between the observed points), by $d\phi/dt$ for the
quiescent galaxies. By construction, this fraction has a typical value
of around unity, as we have set the visibility timescale based on the
assumption that 100\% of the new quiescent galaxies have formed
following a post-starburst phase. We can now see clearly that this
fraction appears to be relatively constant with redshift. While it is
tempting to observe that the contribution of the PSB galaxies appears
to increase with decreasing redshift for the highest mass galaxies,
but decrease for lower mass galaxies, any trend is not significant
within the errors.

We note that, in this discussion, we have neglected the impact of
dry-mergers which could move galaxies that are already quiescent into
higher mass bins and therefore contribute to the apparent growth rate
in a single mass bin.  Should the stellar masses of our quiescent
galaxies be slightly overestimated due to an under-estimation of the
contribution of the TP-AGB phase, the post-starburst contribution at a
given stellar mass would be even higher, leading to a shorter estimate
for the visibility window. As this is unlikely, given our knowledge of
the post-starburst evolutionary phase (and see next section), we
conclude that it is likely that there is no significant bias in our
quiescent galaxy masses. 

\subsection{``Switch-off'' vs. ``starburst'' mechanisms}

\begin{figure}
 \includegraphics[scale=0.55]{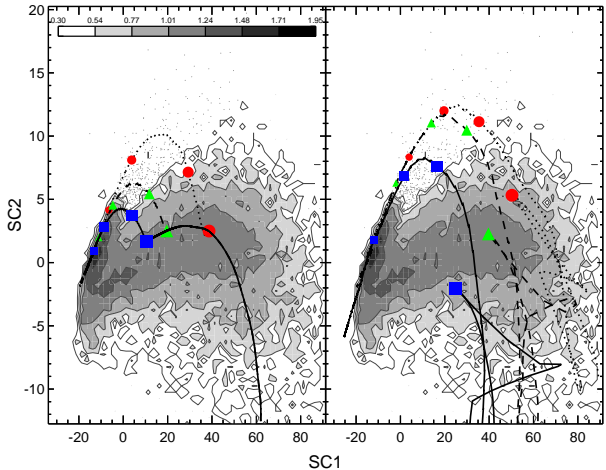}
 \caption{The distribution of the first two super-colours (SC) for
   galaxies with $0.5<z_{\rm phot}<1.0$ observed in the UDS field,
   using a $\log_{10} N$ grey-scale with bin sizes of $\Delta {\rm
     SC1}=2.5$ and $\Delta {\rm SC2}=0.5$. The post-starburst
   population are shown as dots. Overplotted in black are the
   evolutionary tracks of composite solar metallicity stellar
   population models with a dust attenuation of $A_V=1$. {\it Left:}
   solar metallicity models with continuous star formation truncated
   after 1\,Gyr (dotted, red circles), 3\,Gyr (dashed, green
   triangles) and 6\,Gyr (full, blue squares). The star formation is
   truncated exponentially, over a timescale of 50\,Myr. Symbols mark
   time of truncation, then 0.1, 0.5 and 1\,Gyr following truncation,
   with decreasing symbol size. {\it Right: } with a constant star
   formation rate for 1\,Gyr, followed by a burst of star formation
   that accounts for 50\% of the total mass of the galaxy.  The
   starburst is an exponential with a decay time of 50\,Myr.  The
   three tracks are for models with metallicities of 0.25\,$Z_\odot$,$
   Z_\odot$ and 2.5\,$Z_\odot$. Symbols mark the time of the
   starburst, then 0.3, 0.5 and 1\,Gyr after the starburst, with
   decreasing symbol size. } \label{fig:SC12_models}
\end{figure}

We now return to the question of ``switch-off'' vs. ``starburst'' as
two mechanisms for the rapid quenching of star formation in
galaxies. In Figure \ref{fig:SC12_models} we show a greyscale of the
first two super-colours of the observed UDS galaxies for our lowest
redshift bin ($0.5<z_{\rm phot}<1.0$), and overplotted evolutionary
tracks from BC03 spectral synthesis models. In the left panel the
three tracks show the evolution in colour of a solar metallicity
stellar population with a continuous star-formation rate, truncated
after 1\,Gyr (dotted line), 3\,Gyr (dashed line) and 6\,Gyr (full
line). The star formation is truncated exponentially, over a timescale
of 50\,Myr.  The models are attenuated by a moderate amount of dust,
with $A_V=1$ using a 2-component dust attenuation law
\citep{wild_psb,2008MNRAS.388.1595D}, although this does not
significantly impact their position in SC space during the PSB or
quiescent phase. Symbols mark the point of truncation and then 0.1,
0.5, and 1\,Gyr following truncation. We can see that continuous
star-formation histories that are truncated before 6\,Gyr enter the
region of super-colour space that we have defined to be
post-starburst, although can not explain the most extreme
objects. While one could argue that a 1\,Gyr truncated population was
indeed a ``post-starburst'' galaxy, with a rather extended burst
period, this terminology becomes less obvious for the longer
timescales. These more extended star formation histories enter the
``post-starburst'' region of super-colour space for anywhere between a
hundred Myr and nearly a Gyr.

In the right panel of Fig. \ref{fig:SC12_models} we present models
which have undergone 1\,Gyr of continuous star formation, followed by a
starburst that accounts for 50\% of the total stellar mass of the
galaxy. Similar to the previous case, the starburst decays over a
timescale of 50\,Myr, although extending this to 100\,Myr makes little
difference to the evolutionary tracks.  Models with three
metallicities are plotted: 0.25\,$Z_\odot$,$ Z_\odot$ and
2.5\,$Z_\odot$.  Symbols mark the start of the starburst and then 0.3,
0.5, and 1\,Gyr following the starburst. While it is possible to
explain the colours of the majority of the post-starburst galaxies
with solar metallicity, truncation or moderate burst models, the
post-starburst galaxies with the most extreme colours in our samples
may have lower metallicities and very substantial burst mass
fractions. These extreme objects must have a visibility timescale
closer to 1\,Gyr i.e. longer than the $\sim$250\,Myr derived above. If
a significant fraction of the post-starburst galaxies have such a long
visibility window, this allows for some growth of the quiescent
population to take place via a slower quenching route, that would not
be visible to our study.

In reality, the population may have a mixed origin, with some fraction
caused by truncation or weak starbursts, and others caused by more
extreme events. We tested whether the PSB mass functions varied with
strength of SC2, by splitting the samples into two at SC2$=7$, finding
that they did not. Therefore, if a substantial fraction of the
galaxies that lie close to the red sequence are galaxies that have
simply truncated their star formation, or undergone only a weak
starburst, the process that causes this must have an equal efficiency
as a function of mass as processes which cause much stronger
starbursts. 

Ultimately, the strength of the PSB features depends on several
factors: the ratio of old (non-burst) to burst stellar mass in the
galaxy; the star-formation history of the non-burst mass; the rate of
decline of the burst. These factors may be difficult to disentangle
for any one galaxy but could perhaps be constrained in the future
through detailed analysis of the optical spectra and SEDs, and through
population studies.

\subsection{Two formation mechanisms for \boldmath{$z>0.5$}
  post-starburst galaxies}\label{sec:formation} 

It is clear that the mass function of transitioning galaxies should
match in shape those of recently quenched red sequence galaxies
\citep{Peng:2010p9523}. At high redshift we have the advantage of the
short amount of cosmic time that has passed, meaning that all red
sequence galaxies are recently quenched galaxies. It is therefore very
interesting to note that at $z\gsim1.5$, the shape of the PSB mass
function is similar to that of the quiescent galaxy mass function
(Fig.  \ref{fig:MF}). Splitting the sample into smaller redshift bins
confirms the trend that by $z\sim2$ the shapes of the quiescent and
PSB mass functions are almost indistinguishable. Quantitatively, the
number density of the PSB galaxies is a factor of 4 lower than the
quiescent galaxies at these redshifts. On the other hand, the age of
the Universe is $\sim4$\,Gyr, which is a factor of 4 larger than the
maximum likely visibility timescale of the PSB galaxies. This is again
consistent with {\it all} quiescent galaxies already in place by
$z\sim2$ having had their star formation truncated rapidly. The
difference between the shape of the star-forming and PSB
mass-functions at high redshift also tells us that PSB galaxies are
not a randomly drawn sample from the star-forming population. A hidden
factor related to stellar mass must dictate whether a galaxy can be
quenched quickly enough to be detected as a PSB galaxy.

At $z\lsim1$ the mass function of the PSB population becomes
increasingly similar to the mass function of the star-forming
population (lower panel of Fig. \ref{fig:MF}). The strength of the
spectral features required in order to detect PSB galaxies requires a
significant prior specific star-formation rate. Therefore, it makes
sense that the mass function of the PSB galaxies closely resembles
that of the ``progenitor'' population of gas-rich star-forming
galaxies, but this also implies that the dominant mechanism that is
responsible for ``fast-quenching'' is independent of stellar mass at
$0.5<z<1$.

Our quiescent mass function does not show such a strong flattening
with decreasing redshift as seen in previous work. However, as we note
in Section \ref{sec:mf}, if we combine both the PSB and quiescent mass
functions as would happen if we were using a traditional $UVJ$ or
NUV-$r$-$J$ colour selection, the low mass ends of our intermediate
and low redshift quiescent galaxy mass functions do become
flatter. \citet{Peng:2010p9523} predict that the flattening and
associated low-mass upturn is due to galaxies which have been quenched
due to their environment; our results provide a rather tighter
constraint as the flattening appears to originate at least partly and
possibly wholly from \emph{rapidly} quenched galaxies. This could rule
out some cluster-specific processes such as harassment and
strangulation as these may happen too slowly to provide the strength
of features needed for the galaxies to be identified as PSBs. Our
results are consistent with the idea that pre-processing in group
environments, where the velocity dispersions are low enough for galaxy
mergers to occur, is key to building the low-mass end of the red
sequence at $z<1$. These ideas are explored further by studying the
environments of the PSB galaxies in Socolovsky et~al. (in prep.).

To summarise, our results imply a two-phase mechanism for the rapid
quenching of galaxies at $z>0.5$. At high redshifts ($z>2$), quiescent
galaxies form rapidly from the dissipational collapse of gas clouds,
or rapid merging of many smaller proto-galaxies, resulting in intense
star formation for a short period of time, and a single quenching
episode. The fact that PSB galaxies are not randomly selected from the
star-forming galaxy mass function implies that a preferred mass scale
exists, either for the formation or the quenching mechanism (i.e.
``mass-quenching'').  At low redshift ($z<1$), a significant fraction
of quiescent galaxies are quenched rapidly, at all masses, from the
star-forming population. Further studies of the environment and
clustering of the PSB galaxies (Wilkinson et al. in prep.; Socolovsky
et al. in prep.) will study whether these processes are consistent
with ``environment-quenching'' \citep{Peng:2010p9523}.

\subsection{The progenitors of post-starburst galaxies: Sub-mm galaxies?}

In order to detect the strongest PSB features in broad band spectral
energy distributions, it is necessary for the galaxies to have
undergone a recent enhanced period of star formation of relatively
short duration. Their progenitors should easily be visible in our high
redshift fields. Observationally it is difficult to distinguish a
galaxy with a constant high SFR, from one undergoing a burst that is
about to rapidly decline as it depletes its cold gas supply. One rare
type of galaxy which may be undergoing an intense but unsustainable
starburst are the sub-millimetre detected galaxies, which have number
densities of $\sim10^{-5}$Mpc$^{-3}$ at $1<z<2.5$
\citep{Swinbank:2014p9524}.  Stellar masses are difficult to
  estimate, due to the significant dust attenuation,
  \citet{Simpson:2014p9589} suggest typical stellar masses of
  \logM$\sim10.9$ for a Salpeter initial mass function, equivalent to
  \logM$\sim10.6$ for the Chabrier initial mass function assumed in
  this paper. Assuming a 100\,Myr timescale for the star formation
activity of sub-mm galaxies \citet{Simpson:2014p9589} estimated that
their space density at $z\sim0$ would be consistent with local
elliptical galaxies of the same near-infrared luminosity distribution
(see also Dunlop 2001)\nocite{Dunlop:2001p10154}. A more recent study
of the clustering of a much larger sample of sub-mm galaxies in the
Scuba-2 Cosmology Legacy Survey by Wilkinson et al. (submitted),
confirms that the highest redshift ($z>2.5$) sources are consistent
with being the progenitors of local massive elliptical galaxies in
clusters. However, sub-mm galaxies at $1<z<2$ are less strongly
clustered, and inconsistent with being the progenitors of the massive
elliptical galaxies in the local Universe.  
  \citet{Smail:2014p10271} also find that sub-mm galaxies associated
  with a cluster at $z=1.62$ are preferentially located in the
  outskirts of the cluster, and will form less massive ellipticals
  than those galaxies that are already quiescent by this redshift.

With a very similar characteristic stellar mass of
\logM$\sim10.4-10.5$, a space density 6-7 times higher at
$\sim6-7\times10^{-5}$Mpc$^{-3}$ but an observable timescale up to an
order of magnitude longer, the coincidence between the number
densities of the PSB galaxies at $z>1$ and sub-mm galaxies is
suggestive. While it is difficult to determine the star-formation
history of the sub-mm galaxies due to the dominant contribution of the
young, high mass stars to the spectral energy distribution
\citep{Simpson:2014p9589},  and significant dust attenuation,
tighter constraints can be obtained for the PSB galaxies where the
youngest stars have exploded as supernova and thus no longer dominate
the SED.  Although the errors in the comparison are large, the
  coincidence also supports the modelling presented in Figure
  \ref{fig:SC12_models}, which shows that the extreme colours of a
  fraction of the PSB galaxies can only be explained if they have
  undergone an intense and unsustained episode of star formation,
  shortly prior to their detection. Strengthening the connection
between PSB and sub-mm galaxies may therefore help constrain the
physical properties of both populations of galaxies in the future.

\section{Summary}

We have extended the method presented in WA14 to investigate the
number density and mass functions of post-starburst (PSB) galaxies in
the UKIDSS Deep Survey (UDS), and how they relate to the increase in
number density of quiescent galaxies with \logM$>10$ at $0.5<z<2$. Our
main results can be summarised as follows:
\begin{itemize}

\item The shape of the mass function of post-starburst galaxies
  resembles that of quiescent galaxies at $z\sim2$, but by $z\sim0.5$
  more closely resembles that of star-forming galaxies.

\item The evolution in shape of the mass function of the PSB galaxies
  implies a two-phase mechanism for the rapid quenching of galaxies at
  $z>0.5$: (1) rapid dissipational collapse at high redshift, which
  occurs exclusively in the high-mass galaxy population; and (2) rapid
  quenching of star-forming galaxies at low-redshift, independent of
  stellar mass.

\item By adding the mass function of the PSB galaxies to that of the
  quiescent galaxies in the intermediate and high redshift bins, we
  are able to reproduce exactly the shape of the quiescent mass
  function in the subsequent (low and intermediate) redshift
  bins. This shows that 100\% of the growth in the quiescent
  population at $0.5<z<2$ could have arisen following a PSB phase. If
  this is the case, then PSB features must typically be visible for a relatively
  short time of $\sim$250\,Myr.

\item This consistency of true and predicted mass functions shows that
  the flattening in the low-mass end of the quiescent galaxy mass
  function observed in previous work is consistent with originating
  entirely from \emph{rapidly} quenched galaxies.

\item Equivalently, by comparing the rate of increase in the number
  density of quiescent galaxies with cosmic time to the occurance of
  post-starburst galaxies, we show that there is no preferred stellar
  mass or redshift at which the post-starburst galaxies contribute to
  the growth of the red sequence. However, larger surveys are needed
  to fully explore these relations.

\item Comparison with spectral synthesis models show that the
  super-colour criteria are effective for isolating recently-quenched
  galaxies which have formed a significant fraction of their mass
  within the last $\sim$Gyr, and this may include galaxies which have
  ``switched-off'' star formation, as well as genuine
  ``post-starburst'' galaxies. 

\item The galaxies with the most extreme
  colours must have formed a significant fraction of their stellar
  mass in a short-lived starburst, and may have metallicities below
  solar. These most extreme post-starburst galaxies must be visible
  for longer than the 250\,Myr estimated via number density arguments,
  which would leave some space for slower quenching mechanisms to
  contribute to the build up of the quiescent population. 

\item The consistency of the PSB mass functions close to, and further
  away from the quiescent population suggests that there can be no
  difference between the mass efficiency of the physical mechanisms
  that cause ``switch-off'' vs. true ``post-starburst'' galaxies. Further
  analysis of the spectra and SEDs, as well as population studies, may
  help to distinguish between the two scenarios.

\item If the post-starburst spectral features are visible for
  significantly longer than 250\,Myr, or the galaxies rejuvenate their
  star formation and return to the blue-sequence following their
  period of rapid quenching, this allows for an additional
  ``slow-quenching'' mechanism to contribute to the build up of the
  quiescent population. Our results show that the mass efficiency of
  this mechanism must be similar to that of the ``fast-quenching''
  mechanism, in order to reproduce the mass function of the quenched
  galaxies at later times.

\item The stellar masses and number densities of PSB galaxies are
  consistent with them being the descendants of sub-mm galaxies,
  although the errors on this comparison are large. Strengthening the
  connection between these two populations may help to better
  constrain the physical properties of both.

\end{itemize}

Further work on this very interesting class of recently quenched
high-redshift galaxies is ongoing. However, their transitory nature
means that ultimately larger area and deeper surveys will be needed to
fully quantify the role of different mechanisms in the origin of
galaxy bimodality and build up of the red-sequence.





\section*{Acknowledgements}
UKIDSS uses the UKIRT Wide Field Camera
\citep[WFCAM][]{2007A&A...467..777C}. The photometric system is
described in \citet{2006MNRAS.367..454H}, the calibration is described
in \citet{Hodgkin:2009p4949} and the science archive is described in
\citet{2008MNRAS.384..637H}. We are indebted to the staff at UKIRT for
their tireless efforts in the face of very difficult circumstances.

We would like to thank the referee of the paper for the constructive
and helpful comments. V.~W. and K.~R. acknowledge support of the
European Research Council via the award of a starting grant (SEDMorph;
P.I. V.~Wild). J.~S.~D. acknowledges support from the European
Research Council via the award of an advanced grant
(P.I. J. Dunlop). R.~J.~M. acknowledges the support of the European
Research Council via the award of a consolidator grant
(P.I. R. McLure) J.~S.~D. also acknowledges the contribution of the EC
FP7 SPACE project ASTRODEEP (Ref.No: 312725).

\bibliographystyle{mn2e}


\end{document}